\documentclass{ws-procs975x65}

\usepackage{bm}
\usepackage{color}

\newcommand{\be}{\begin{equation}}
\newcommand{\ee}{\end{equation}}
\newcommand{\bea}{\begin{eqnarray}}
\newcommand{\eea}{\end{eqnarray}}

\newcommand{\benu}{\begin{enumerate}}
\newcommand{\eenu}{\end{enumerate}}
\newcommand{\bite}{\begin{itemize}}
\newcommand{\eite}{\end{itemize}}

\begin{document}

\title{Application of Superstatistics to  atmospheric turbulence}
\author{SALVO RIZZO }
\address{E.N.A.V. S.p.A., U.A.A.V. Firenze, Italy\\e-mail:salvrizzo@tiscali.it}
\author{ANDREA RAPISARDA }
\address{Dipartimento di Fisica e Astronomia Universit\`a di Catania, and \\
INFN sezione di Catania, CACTUS Group, 
\\Via S. Sofia 64, 95123 Catania, Italy\\e-mail:andrea.rapisarda@ct.infn.it}

\maketitle

\abstracts{We successfully apply the recent developed
superstatistics theory to a temporal series  of turbulent wind
measurements recorded by the anemometers of  Florence airport.
Within this approach we can reproduce very well the fluctuations
and the pdfs of wind  velocity returns and differences.}

\section{Introduction}

During the last decades an enormous effort has been devoted to
understand the physical origin of turbulence, both at the
theoretical level and at the experimental
one\cite{kolm,obu,krai,cast,pope,fri,sree}.
Although
many steps forward have been done,
 a well established theory of turbulence does not yet exist and many
fundamental aspects of this phenomenon remain still unclear.

 Atmospheric turbulence is  a challenge  \textit{per se} being characterized
 by very high Reynolds numbers
 ($Re\sim 10^8$) and  very intermittent distributions. On the other hand
 beyond basic research its interest lies also on the several engineering and
 meteorological
 applications.
The Kolmogorov hypotheses were  verified in the sixties doing
measurements  in the atmospheric boundary layer considering the
flow velocity for a relative short time, with a high sampling rate
and for relatively constant mean flow, in order to control and
maintain constant  the Reynolds number. In general this is not
always possible.


 Although atmospheric turbulence\cite{sree,ramos,peinke} can have peculiar
 features regarding the non-stationarity character of the wind data and the high
turbulence intensity
\footnote{The turbulence intensity of the wind speed are typically
expressed in terms of standard deviation, $\sigma_v$, of velocity
fluctuations measured over 10 to 60 minutes normalized by the mean
wind speed $V$. $[I_v=\frac{\sigma_v}{V}]$. Typical values for
complex terrain are $I_v\geq 0.2$ while, on the other hand for
microscale turbulence one usually has $ I_v\sim O(10^{-2})$}
 many similarities with microscopic turbulence
exists. For a detailed comparison between the two, one can see the
recent paper by B\"ottcher  et al.\cite{peinke}.

In this short paper we discuss  a study
 of a turbulent wind data series recently measured at
 Florence airport for a period of six months. We  show by means of a statistical
analysis that we can describe this  example of atmospheric
turbulence by means of the nonextensive approach adopted in refs.
\cite{ramos,beck1,beck-swin,swin1,tsa1,tsa2}, within the more
general   superstatistics formalism introduced in ref.
\cite{be-co}. The latter justifies the successful application of
Tsallis statistics in different fields, and more specifically  in
turbulence
experiments\cite{beck-swin,swin1,be-co,tsa1,tsa2,vinas1}. We will
show that such an approach is meaningful and can reveal very
interesting features which could have also a very practical
utility for safety reasons when applied to air traffic control
services. Part of this study has just been published \cite{rizzo1}
and a longer paper with a complete and exhaustive analysis is in
preparation \cite{rizzo2}.

\section{Statistical analysis of wind measurements }

 The wind velocity  measurements,
were  taken at  Florence  airport and  were  done for  a time
interval  of six months, from October 2002 to March 2003. Data
were recorded by  using two 3-cups runway heads anemometers, each
one mounted on a $10$ $m$ high pole, located at a distance of 900
$m$ and with a sampling frequency of one measure  every 5 minutes.
Although in our experiment  we actually could not  control the
Reynolds number, as usually done in microscopic turbulence, and
despite our low sampling frequency  ($3.3\cdot10^{-3}$Hz) and the
high intermittency of our wind data, we  found several features of
{\it canonical} turbulence as we will discuss  in the following.
We performed, on our time series, a statistical analysis using
conventional mathematical tools which are normally adopted  in
small scale physical turbulence studied in laboratory. In
particular we  investigated   correlations,  spectral
distributions as well as  probability density functions of
velocity components of returns  and differences see refs.
\cite{rizzo1,rizzo2} for more details. In this short contribution
for simplicity and lack of space we discuss only returns of the
longitudinal velocity components measured by one of the two
anemometers (in the present case   the one closest to the runway
head 05 and labeled {\em RWY05}) defined by the following
expression
\be \label{eq1}
 x(t)_{\tau} = V^{RWY05}_x(t+\tau) - V^{RWY05}_x(t) ~~~~,
\ee
\noindent $V_x(t)$ being the longitudinal velocity component at
time $t$ and $\tau$ being a fixed time interval \footnote{Returns
are here defined in a slight different way from those used in
econophysics, i.e.: $\frac{x(t+\tau)-x(t)}{x(t)}$. }. The same
analysis was done also for the transversal components and for
velocity difference between the two anemometers with similar
results \cite{rizzo1,rizzo2}.

\subsection{ Correlations and power spectra }

Our data show very strong correlations and  power spectra with the
characteristic -5/3 law in the high-mid portion of the entire
spectrum, see fig.1 in ref.\cite{rizzo1}.  However the dissipation
branch in the high-range frequency, well known in micro-scale (or
high-frequency) turbulence analysis \cite{kolm,fri}, is here
missing due to the low-frequency sampling used
\cite{rizzo1,rizzo2}. Correlation functions also show an initial
exponential decay, followed by a power law-decay modulated by the
day-night wind periodicity which is a well known phenomenon. No
significant difference was found for day and night periods, when
air traffic is almost absent. For more details please see refs.
\cite{rizzo1,rizzo2}.

\subsection{The superstatistics approach for wind velocity pdfs}

The superstatistics formalism proposed recently by C. Beck and
E.G.D. Cohen is a general and effective description for
nonequilibrium systems. For more details see the original article
 and the paper by Beck in this volume\cite{be-co}.
 In the superstatistics approach one considers fluctuations of an
intensive quantity, for example the temperature, by introducing an
effective Boltzmann factor
 \be
   B(E)= \int_0^\infty f(\beta) e^{-\beta E} d\beta
\ee where $f(\beta)$ is the probability distribution of the
fluctuating variable $\beta$, so that we have for the probability
distribution \be P(E)= \frac{1}{Z} B(E)~~, \ee with the
normalization given by \be Z=\int_0^\infty B(E) dE~~~. \ee
 One can imagine  a collection of many cells, each of one with
 a defined intensive quantity, in which a test particle is
 moving. In our atmospheric turbulence studies, the time series
 of the wind velocity recordings,
 are characterized by a fluctuating variance, so the returns
 (\ref{eq1}), cannot be assumed to be  by a "simple" Gaussian process.
 They show a very  high intermittent behavior stronger than  that  one
 usually found in small-scale fluid turbulence experiments.
\begin{figure} 
\centerline{ \epsfxsize=3.2in\epsfbox{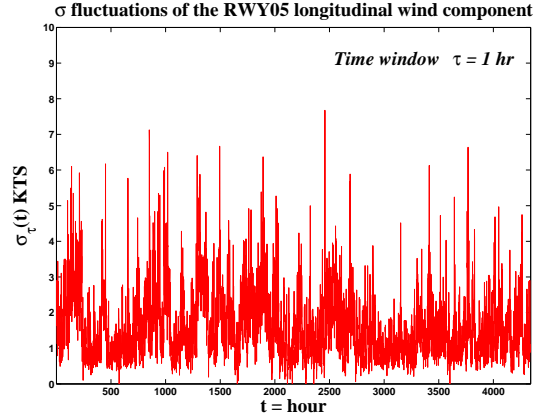}}
\caption{Variance fluctuations of the longitudinal wind velocity
component for the anemometer RWY05 obtained with a moving time
window $\tau$ of one hour.}
\end{figure}
\begin{figure}
\centerline{ \epsfxsize=3.2in\epsfbox{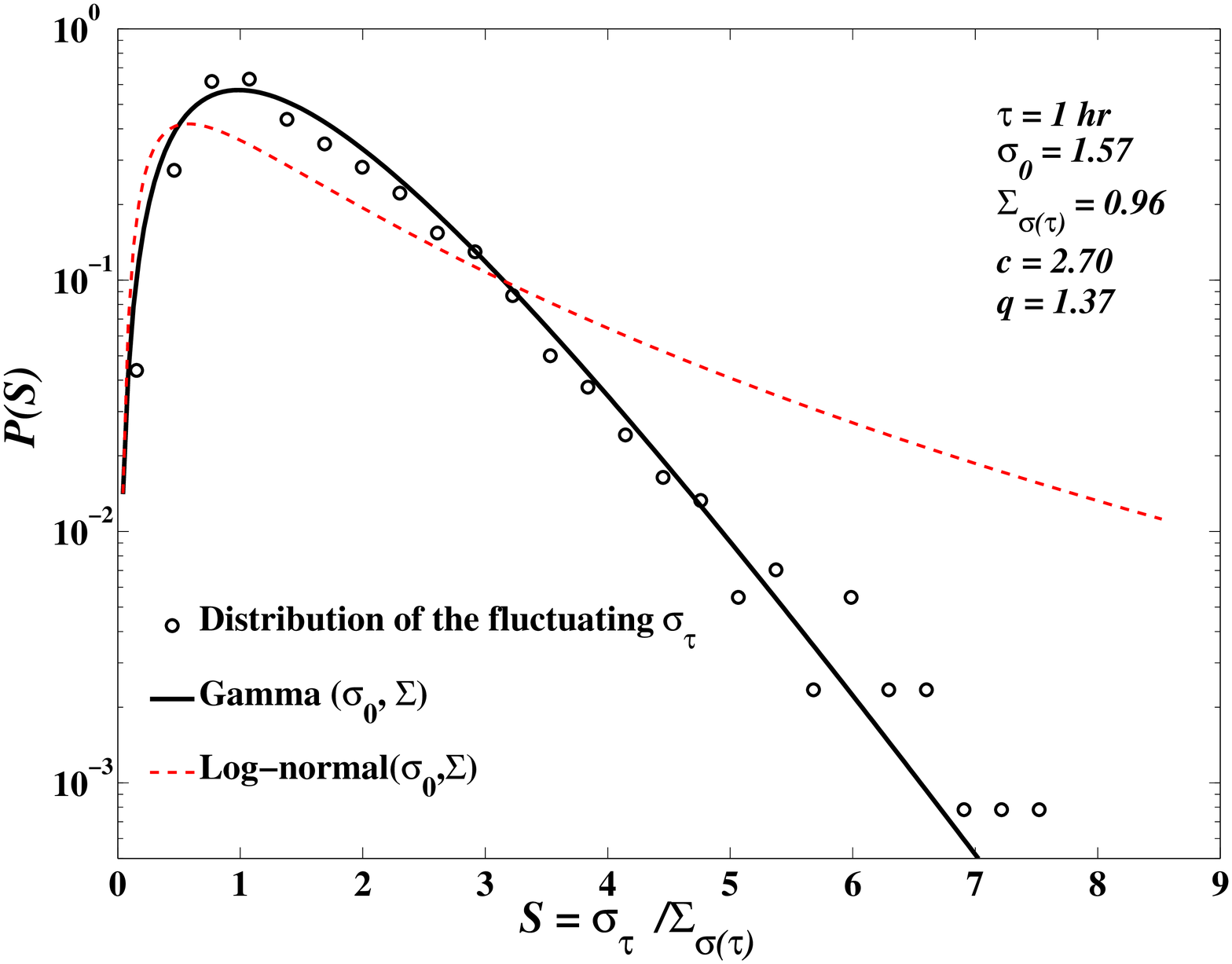}}
 \caption{Standardized pdf of the fluctuating variance
corresponding to the previous figure (open points) are compared
with a Gamma distribution (full line) and with a Log-normal
distribution (dashed line) sharing the same mean ($\sigma_0$) and
variance ($\Sigma$) extracted from experimental data. The
Log-normal is not able to reproduce the experimental data.}
\end{figure}
\begin{figure}
\centerline{ \epsfxsize=3.2in\epsfbox{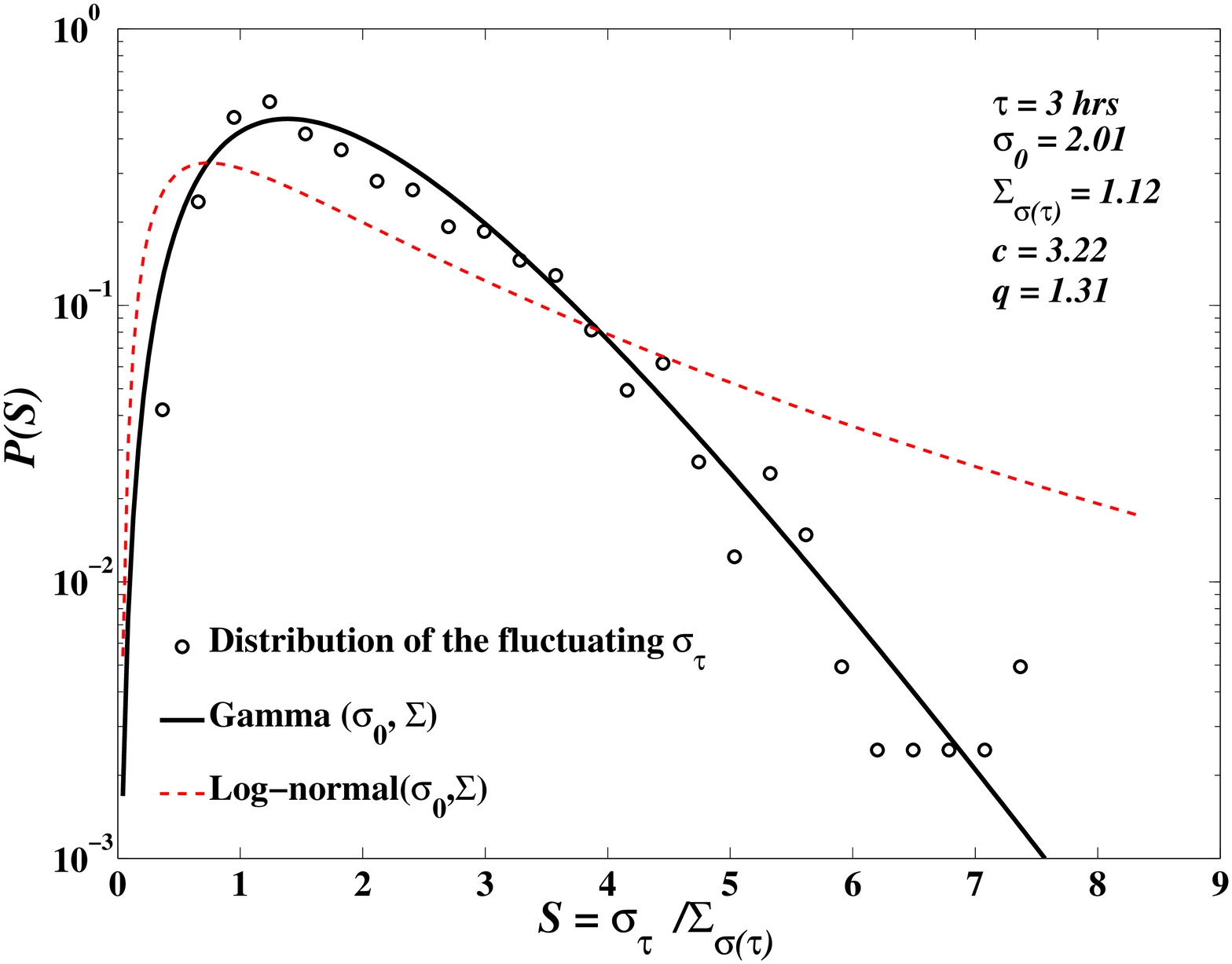}}
\caption{Standardized pdf of the fluctuating variance similar to
the previous figure but  corresponding to a windowing of three
hours in our longitudinal velocity components (open points). We
show for comparison also a Gamma distribution (full line) and a
Log-normal distribution (dashed line) sharing the same mean
($\sigma_0$) and variance ($\Sigma$). Also in this case the  Gamma
distribution reproduces very well the experimental data at
variance with the Log-normal one.}
\end{figure}
%


In our analysis  we considered the following quantities:
\emph{(a)} the wind velocity returns $x$ defined by
eq.(\ref{eq1}), \emph{(b)} the corresponding  variance of the
returns $x$ which we indicate with $\sigma $, \emph{(c)} the
fluctuations of $\sigma $, whose variance we indicate with the
symbol $\Sigma$.

We extracted from the experimental data, using an fixed time
interval $\tau$, the distribution for the fluctuations of the
longitudinal wind component variance. The aim is to slice the time
series in "small" pieces in which the signal is almost Gaussian
and apply  superstatistics theory. This fluctuating behavior of
$\sigma$ is plotted in Fig.1 for a time interval $\tau= 1$ hour.
In Figs. 2 and 3 we then plot the probability distribution of the
variance $\sigma$ for $\tau=1$ and $3$ hours respectively. In the
same figures we plot for comparison a Gamma (full curve) and a
Log-normal (dashed curve) distribution\cite{fri} characterized by
the same average and variance extracted from the experimental
data. In this sense, the curves are not fitted to  the data.
The comparison clearly  shows that  the  Gamma distribution is
able to reproduce very nicely the experimental distribution of the
$\sigma$ fluctuations and that this type of distribution show
robustness for different time windows choices.   This is at
variance with the Log-normal distribution which is usually adopted
in microscopic turbulence and which in this case is not able to
reproduce the experimental data.

In general using Beck and Cohen notation\cite{be-co} we have for
the Gamma distribution \be f(\beta) = \frac{1}{b \Gamma(c)} \left(
{ \frac{\beta}{b}} \right) ^{c-1} e^{-\beta/b} \ee with \be
  c=\left( {\frac{\sigma(\beta)}{b} } \right)^2 =  \frac{1}{q-1} ~~,  ~~~~~~~
  bc=<\beta>=\beta_0
\ee where $2c$ is the actual number of effective degrees of
freedom and $b$ is a related parameter.
Inserting  this distribution into the generalized Boltzmann factor
(2) one gets the q-exponential curve \cite{tsa1},
\be
         P(x)= \left(1-(1-q) \beta_0 E(x) \right)^{\frac{1}{1-q}} ~~.
\ee
In our analysis we have   $E=\frac{1}{2}x^2$ with $x$ defined by
eq.(1)\cite{beck-swin,rizzo1,rizzo2}.
Considering  the fluctuations of the variance $\sigma$ of the
returns  $x$, we get  the following correspondence with the
original superstatistics formalism
\be
  \beta= \sigma_{\tau} ~~~,~~~~~
  \sigma(\beta)= \Sigma({\sigma_{\tau}})~~~,~~~~~
  \beta_0=<\sigma_{\tau}>=\sigma_{0}~~~.
\ee
%
In the present case, we get for the Gamma distributions  which
describe the experimental variance fluctuations reported in Figs.2
and 3 the characteristic values $c=2.70$ and  $c=3.22$,  for a
time interval of $1$ and $3$ hours, from  which, using eq. (6), we
get the corresponding q-values $q=1.37$ and  $q=1.31$.
 In Fig.4 we plot the probability density function $P(x)$
of the experimental longitudinal returns for different time
intervals, i.e.  1 hour (full circles), 3 hours (open diamonds)
and 24 hours (open squares). For comparison we plot a Gaussian
distribution (dashed curve) and the q-exponential curves (7)
 characterized by the q-values extracted from the Gamma
 distributions of Figs.2 and 3 for a time interval $\tau$
 corresponding to $1$ and $3$ hours. 
 The q-exponential curves reproduce very well the experimental
 data which, on the other hand, are very different from the Gaussian pdf. However one can
 notice that for a very long time interval, i.e. $\tau=$24 hours,
 the data are not so far from being  completely decorrelated and
 therefore the corresponding experimental pdf is closer to the
 Gaussian curve. Notice that
  the theoretical curves are not fitted, and that the superstatistic approach,
  in a self-consistent and elegant way,  is able to
  explain and characterize in a quantitative way
  the wind data.
In a similar way one  can extract theoretical curves
  which reproduce
 the wind velocity differences pdfs with similar entropic  q-values,
 although in that case an asymmetry correction
 has to be considered to better reproduce the tails of the
 pdfs\cite{rizzo1,rizzo2}.

\begin{figure}
\centerline{\epsfxsize=3.20in\epsfbox{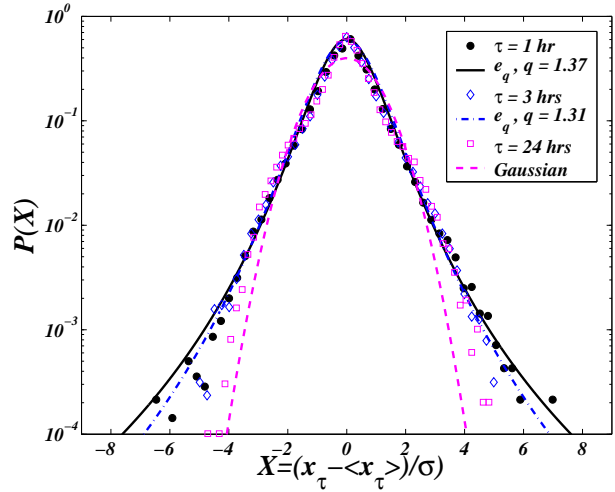}}
\caption{Comparison between standardized longitudinal velocity
returns pdfs for three different time intervals ($\tau=1,3,24$
hours)and the q-exponential curves with the q-value extracted from
the $c$ parameter of the Gamma distribution  shown in the previous
figures. A Gaussian pdf is also shown as dashed curve. See text.}
\end{figure}
 In our analysis the
large and intermittent  wind velocity variance fluctuations are
reproduced very well by  a Gamma type superstatistics excluding
the Log-normal one and this gives exactly the Tsallis
q-exponential for the velocity returns pdfs. However one has to
say that the situation is much more difficult for the less
fluctuating velocity flow of the microscale fluid turbulence.


We add as a  final remark that, very recently a similar method has
been adopted by a research group at  the NASA Goddard Space Flight
Center to analyze the solar wind speed fluctuations\cite{vinas1}.

\section{ Conclusions }

 We have studied a temporal series of wind velocity
 measurements recorded at
Florence airport for a period of six months. The statistical
analysis for the velocity components shows intermittent
fluctuations which exhibit power-law pdfs. Applying the
superstatistics formalism, it is possible to extract a Gamma
distribution from the probability distributions of the variance
fluctuations of wind data. The  characteristic parameter $c$ of
this Gamma distribution gives the entropic index $q$ of the
Tsallis q-exponential which is then able to reproduce very well
the velocity returns and differences pdfs. Beyond the successful
application of superstatistics and Tsallis thermostatistics  for
turbulent phenomena and the corresponding theoretical
implications, we think that this work shows  a useful and
interesting method to characterize and study in a rigorous and
quantitative way atmospheric wind data  for safety flight
conditions in civil and military aviation.

\section*{Acknowledgements}

The authors are indebted to   C. Beck, E.G.D. Cohen, S. Ruffo,
H.L. Swinney and  C. Tsallis for suggestions and discussions.

\end{document}